\documentclass[12pt,reqno]{amsart}
\usepackage{amsmath,a4wide}
\usepackage{xfrac}
\usepackage{stmaryrd,mathrsfs,bm,amsthm,mathtools,yfonts,amssymb,color}
\usepackage{esint}
\usepackage{xcolor}
\usepackage[pagebackref,citecolor=black,linkcolor=blue,colorlinks=true]{hyperref}
\usepackage{tikz,tikz-3dplot}

\begingroup
\newtheorem{theorem}{Theorem}[section]

\endgroup

\theoremstyle{definition}

\title{Weak stablity and closure in turbulence}

\author[C. De Lellis]{Camillo De Lellis}
\address{School of Mathematics, Institute for Advanced Study, 1 Einstein Dr., Princeton NJ 05840, USA}
\email{camillo.delellis@math.ias.edu}
\author[L. Sz\'ekelyhidi, Jr.]{L\'aszl\'o Sz\'ekelyhidi, Jr.}
\address{Institute of Mathematics, University of Leipzig, Augustusplatz 10, Leipzig 04109, Germany}
\email{laszlo.szekelyhidi@math.uni-leipzig.de}

\begin{document}

\begin{abstract}
We survey recent results in the mathematical literature on the equations of incompressible fluid dynamics, highlighting common themes and how they might contribute to the understanding of some phenomena in the theory of fully developed turbulence. 
\end{abstract}

\maketitle

\section{Introduction}
A central issue in any theory of turbulence is how to define, calculate or model Reynolds stresses. Of course there is general agreement that Reynolds stresses arise because products of averages are in general not equal to averages of products. However, the notion of taking averages is difficult to formalize mathematically in a useful way: whether long-time averages as proposed by Reynolds or ensemble averages as in Kolmogorov's work, since very little is known about the object to which the averaging is applied (global-in-time solutions of Navier-Stokes or a physically meaningful measure over ensembles), it is very difficult to deduce first principles information beyond dimensional analysis.   

In this paper our aim is to propose an approach to this difficulty, based on weak convergence techniques combined with the method of convex integration. Although a deterministic approach towards turbulence via weak convergence techniques is not at all new, our contribution here is to emphasize two strands of development in the mathematical analysis of nonlinear PDEs which we believe should deserve more attention in fluid dynamics. Firstly, the concept of weak stability of an equation (that is, stability/continuity of certain nonlinearities under weak convergence), when combined with natural energy bounds, may give nontrivial constraints on possible closure models which cannot be easily deduced from ensemble or long-time averages. Secondly, using the method of convex integration one can construct examples of weak solutions which may be

used to show sharpness of the constraints thus obtained as well as sharpness of various scaling laws. We illustrate these points of view with several recent examples. 

\subsection{Acknowledgments} CDL has been supported by the National Science Foundation under Grant No. DMS-1946175. LS has been supported by the European Research  Council under the European Union’s Horizon 2020 research and innovation programme (grant agreement No.724298-DIFFINCL).




\section{Weak convergence as a mathematical tool of averaging}
\label{s:averaging}

In a nutshell, the challenge in turbulence is to be able to efficiently relate macroscopic quantities to the microscopic description of the flow dynamics, a problem shared with many other branches of mathematics and physics. A great difficulty in turbulent flows, however, is that typically there is a large number of active length scales, in contrast with, say, classical homogenization or kinetic theory - hence one refers to \emph{molar} rather than \emph{molecular} scales \cite{LumleyBook}. This lack of scale separation limits the applicability of methods based on formal asymptotic expansion. Furthermore, it leads to difficulties when trying to define averages: both the classical approach of O.~Reynolds via long-time averages and the more modern statistical approach via ensemble averages suffer from this difficulty. 

Starting with the work of A.~Kolmogorov, the probabilistic approach, postulating the existence of a probability measure with certain symmetries, has been very successful in deducing scaling laws, most prominently the $5/3$ law for energy distribution in the inertial range \cite{FrischBook}. However, without further information on how this measure arises, it is difficult to go beyond. There have been many attempts to prove existence of a physically plausible stationary measure, and even though these were very successful in the 2D case (cf. \cite{KuksinBook,HG}), to our knowledge in 3D none has so far succeeded. 

The main argument in favour of a probabilistic description of turbulence is the observed lack of uniqueness \cite{Neumann} combined with the observed reproducibility of statistics. As far as the former is concerned, in the past decade a number of analytical results appeared, showing that even in the deterministic case a high level of non-uniqueness is present in suitable classes of weak solutions. It is worth emphasizing that the non-uniqueness in such examples is not a mathematical pathology, but seems to be a generic phenomenon, strongly suggesting that a probability measure on ensembles with restored symmetries may exist even without having to resort to stochastic modifications of the basic continuum equations. In the next sections we will survey some of these results. For the moment we wish to point out that such results originate from a deeper understanding of the interaction between micro- and macro-scales, which is obtained from applying techniques from two different strands of development in mathematical analysis: on the one hand weak convergence techniques, in particular the point of view pioneered by L.Tartar (cf. \cite{Tartar79}) of studying special nonlinear quantities stable under weak convergence, and on the other hand the programme of M.~Gromov (cf. \cite{Gromov}) of viewing systems of partial differential equations as differential inclusions together with the tools he developed to solve them, first and foremost convex integration.

The idea to use weak convergence techniques to obtain a deterministic alternative to probabilistic theories of turbulence is not new. J.~Leray in his landmark paper \cite{Leray} not only introduced the notion of weak solution to the Navier-Stokes equations on $\mathbb R^3$ which includes the energy inequality (which he referred to as \emph{solutions turbulentes}), but proved its global existence by using weak convergence and compactness. 

Concerning the inviscid limit, a programme to characterize possible oscillations and concentrations in the velocity field has been initiated by DiPerna and Majda \cite{DM2}, with considerable success in the 2D case, in particular with the proof of global existence of weak solutions to vortex sheet initial data by J.~M.~Delort \cite{Delort}. More generally, it has been proposed by P.~Lax in \cite{Lax} that ensemble averages could be replaced by weak limits, in analogy with the zero dispersion limit of KdV, although the author did point out that an important difference to hydrodynamical turbulence is that oscillations in the dispersive context appear on a single scale. This approach was further investigated in Bardos et al \cite{Bardos} using the Wigner transform. 

The starting point in all these approaches is the formula
\begin{equation}\label{e:R}
    \underset{k\to\infty}{\textrm{w-lim }}u_k=:\bar{u},\quad \underset{k\to\infty}{\textrm{w-lim }}(u_k\otimes u_k)=\bar{u}\otimes\bar{u}+R,
\end{equation}
where $\textrm{w-lim }$ refers to a weak limit in a suitable space, and $R$ can be interpreted as a deterministic analogue of the Reynolds stress tensor. The first task is then to deduce nontrivial information on $R$, such as does it satisfy an equation (c.f.~closure problem) or an inequality? An important point is that in typical situations (such as the one considered by DiPerna and Majda) the approximating sequence $u_k$ is not just weakly convergent but, more importantly, a solution of the Navier-Stokes equations with viscosity tending to zero. 

\subsection{Tartar's framework}

The systematic analysis of possible oscillatory behaviour under a differential constraint has been initiated in the work of L.~Tartar in the 1970s \cite{Tartar79, Tartar05}. The basic framework consists of a vector-valued state variable $z:U\to\mathbb R^N$ defined on an open set $U\subset\mathbb R^d$, which is subject to a set of balance laws
\begin{equation}\label{e:BL}
    \sum_{i=1}^d A^{i}\partial_iz=0\quad\textrm{ in }U,
\end{equation}
and a set of constitutive equations, which may be represented as a submanifold $K\subset\mathbb R^d$ in state-space:
\begin{equation}\label{e:CE}
    z(y)\in K,\quad \textrm{ for all }\,y\in U.
\end{equation}
The central question is to characterize the set of values that can be reached by limits of weakly convergent sequences $z_k$ of exact or approximate solutions. That is, characterize the smallest set $K'$ containing $K$ such that, whenever $z_k$ is a sequence of exact or approximate solutions with $z_k\rightharpoonup \bar{z}$, then $\bar{z}(y)\in K'$ for almost every $y\in U$. 

In this generality we are being vague on purpose concerning the functional setting, e.g.~the precise notion of ``approximate solution'' or the precise notion of ``weak convergence''. The precise setting may vary from case to case (as examples below show), but the general principle is always the same: \emph{characterizing $K'$ amounts to a mathematical closure problem}, which is free from any modelling \emph{ansatz}. Indeed, on an abstract level this can be explained as follows: let $\mathcal{X}$ be a Banach space, representing the space of functions respecting certain \emph{a priori} bounds (e.g.~energy bounds), and let $\mathcal{S}\subset\mathcal{X}$ be the space of solutions of the system under consideration. The main questions are: Is $\mathcal{S}$ weakly closed in $\mathcal{X}$? If not, what is its weak closure $\overline{\mathcal{S}}\subset \mathcal X$? The basic principle behind these questions is that
\begin{center}
\emph{PRINCIPLE I: It is the weak closure $\overline{\mathcal{S}}$ rather than $\mathcal{S}$ which bears physical significance.}    
\end{center} 
One very important mathematical application is the following: if $\mathcal{S}$ is weakly closed in $\mathcal{X}$, there is a good chance that the existence of global solutions can be proved using a suitable approximation scheme. More generally, the same is true for global weak solutions if $\overline{\mathcal S}$ consists of \emph{weak} solutions. Without being too precise, for weak solutions the balance laws and constitutive equations should still be valid in a weak sense (e.g.~\eqref{e:BL} in the sense of distributions and \eqref{e:CE} for almost every $y$).

\subsection{Differential inclusions}

A second strand of development which is relevant to our considerations concerns the theory of (partial) differential inclusions, the simplest type of which can be stated as
\begin{equation}\label{e:DI}
Df(y)\in K,\quad \textrm{ a.e.}\,y\in U.
\end{equation}
Here $U\subset\mathbb R^d$ is an open set as above and $f:U\to\mathbb R^m$ is a (weakly) differentiable map with total derivative $Df(y)$. Typically one considers the boundary value problem, when \eqref{e:DI} is coupled with a condition of the type $f=f_0$ on $\partial U$ for some given $f_0$. 

Early results treated the case when either the domain or the range is one-dimensional \cite{Cellina,CellinaAview,BressanFlores}. Here the key concept is convexity; A typical result is that for linear boundary maps $f_0(x)=a\cdot x$ the differential inclusion \eqref{e:DI} is solvable provided $a\in \textrm{int }K^{co}$, denoting the topological interior of the convex hull. Conversely, it is not difficult to see by integrating that $a\in K^{co}$ is a necessary condition for solvability. In fact more is true: one can show using the Baire category theorem that \emph{typical} solutions of the \emph{relaxed} differential inclusion 
\begin{equation}\label{e:DI-2}
Df(y)\in \textrm{int }K^{co},\quad \textrm{ a.e.}\,y\in U,
\end{equation}
are in fact solutions of the original problem \eqref{e:DI}. 

In the vectorial case, when $d,m>1$, such a general statement cannot be expected, as simple examples demonstrate:
\begin{enumerate}
    \item If $K=O(d)$, the boundary value problem has a solution for any linear boundary map $f_0(x)=Ax$ with $A\in \textrm{int }O(d)^{co}=\{A:\textrm{Id}-A^TA> 0\}$, and, more generally, for any strictly short boundary map $f_0$ - here \emph{strictly short} means that $f_0$ maps any curve to a stricty shorter curve. In analytic terms this is equivalent to the requirement $Df_0(x)\in \textrm{int }O(d)^{co}$ for a.e. $x$. 
    \item If $K=SO(d)$, the only solutions of \eqref{e:DI} are affine maps $f(x)=b+Ax$ with $A\in SO(d)$.   
\end{enumerate}
Thus, in general the geometry of the set $K$ strongly interacts with the differential constraint (to be a gradient) - very much akin to the type of situations studied in \eqref{e:BL}-\eqref{e:CE}. A kind of meta-theorem in the theory of differential inclusions \cite{DacorognaMarcellini, MS99, Kirchheim} is that, up to technical conditions, for existence theorems of \eqref{e:DI} the relaxation $K'$ introduced in connection with \eqref{e:BL}-\eqref{e:CE} can be used as a replacement of the convex hull. However, there is an important technical caveat: the set $K'$ has to be \emph{sufficiently large} in a topological sense. A sufficient condition is that $K'\subset \mathbb R^d$ has nonempty interior, but a key point in Tartar's framework is that certain (linear or nonlinear) constraints may restrict $K'$ to a lower-dimensional variety - a concrete example will be presented below in the context of MHD turbulence. 

Combining the general framework of Tartar in \eqref{e:BL}-\eqref{e:CE} with techniques developed for the existence theory of vectorial differential inclusions \eqref{e:DI} leads to the following basic principle (again formulated in an imprecise way, ignoring technical side conditions):
\begin{center}
    \emph{PRINCIPLE II: Typical elements of $\overline{\mathcal{S}}$ are in fact elements of $\mathcal{S}$.} 
\end{center} 
Here again ``typical'' refers to Baire category. 
As such, using this principle to prove existence of solutions invariably leads to weak solutions with often paradoxical properties which are highly irregular. In this sense this principle is closely related to the construction of Weierstrass' famous example of continuous but nowhere differentiable functions, which can be achieved either by a direct construction of its slowly decaying Fourier series, or by using Baire category. It turns out that both techniques have an appropriate analogy for the general case arising in \eqref{e:BL}-\eqref{e:CE} - the former is often referred to as \emph{convex integration} \cite{MS99}, the latter has been treated in \cite{DacorognaMarcellini,Kirchheim}.

In the past decade the two principles formulated above have led to a new paradigm in fluid mechanics, with the appearance of a number of counterintuitive statements such as the existence of paradoxical weak solutions on the one hand, but also theorems which bear considerable relevance to the physics of fluids and in particular turbulence.   

\section{Examples}

\subsection{The Euler equations} 

Returning to the basic formula \eqref{e:R}, let us now consider the case that the approximating sequence $(u_k)$ consists of solutions of the incompressible Euler equations. That is 
\begin{equation}\label{e:Euler}
\left\{
\begin{array}{ll}
&\partial_t u_k + {\rm div}\, u_k\otimes u_k + \nabla p_k = 0\\ \\
&{\rm div}\, u_k =0
\end{array}\right.
\end{equation}
In addition, since for smooth solutions of the Euler equations the kinetic energy  
$$
\mathcal{E}(u)=\frac{1}{2}\int_{\mathbb T^3}|u|^2\,dx
$$
is a conserved quantity, it is natural to assume a uniform bound of the type
\begin{equation}\label{e:energy1}
    \sup_{t\leq T}\|u_k\|_{L^2(\mathbb T^3)}\leq M
\end{equation}
for some fixed $M$. Then characterizing the weak limit $\bar{u}$ amounts to looking at the weak $L^2$ closure of weak solutions of the Euler equations $\mathcal S$ in the space $\mathcal{X}\subset L^2(\mathbb T^3\times[0,T])$ of solenoidal velocity fields with the energy bound \eqref{e:energy1}. 

As already pointed out in the previous section,
any limit $\bar{u}$ certainly satisfies an equation of the form
\begin{equation}\label{e:Euler-Reynolds}
\left\{
\begin{array}{ll}
&\partial_t \bar{u} + {\rm div}\, (\bar{u}\otimes \bar{u}+R) + \nabla \bar{p} = 0\\ \\
&{\rm div}\, \bar{u} =0
\end{array}\right.
\end{equation}
where $R$ is a deterministic analogue of the Reynolds stress tensor \eqref{e:R}. The ``convexity restriction'' alluded in the previous section leads to the information that, for any $x,t$ the tensor $R(x,t)$ is necessarily symmetric, positive semi-definite and $|\bar{u}|+\textrm{tr }R\leq M$. 

In \cite{DS1,DS2} it is shown that this is in fact the {\em only} constraint satisfied by the weak limit. The main result can be formulated as follows:

\begin{theorem}\label{t:Euler1}
Let $(\bar{u},R)$ be a bounded weak solution of \eqref{e:Euler-Reynolds} such that $R$ is symmetric and pointwise either positive definite or zero. Then there exists a sequence of weak solutions $u_k$ of \eqref{e:Euler} such that $|u|^2=|\bar{u}|^2+\textrm{tr }R$ for almost every $(x,t)$, and $u_k\rightharpoonup \bar{u}$ weakly in $L^2$ as $k\to \infty$.   
\end{theorem}

In the special case when $R$ is a multiple of the identity matrix, $R=\chi\textrm{Id}$, with $\chi$ compactly supported in space-time, this theorem recovers the pioneering result of Scheffer  \cite{Scheffer93}, that is, the existence of  nontrivial weak solutions of the incompressible Euler equations that have compact support in space and time.

Surprisingly, this counter-intuitive ``softness'' of weak solutions of the incompressible Euler equations is common to a number of other equations and problems in fluid dynamics and the techniques used to prove Theorem \ref{t:Euler1} have proved to be surprisingly robust (see \cite{BSV,ChDK,CFG,Shvydkoy,SzVortex,Szekelyhidi,SzWie}

\subsection{The Navier-Stokes equations}
\label{s:NavierStokes}

Proceeding as in the previous section, the natural energy bound for solutions of the Navier-Stokes equations 
\begin{equation}\label{e:Navier-Stokes}
\left\{
\begin{array}{ll}
&\partial_t u_k + {\rm div}\, u_k\otimes u_k + \nabla p_k = \Delta u_k\\ \\
&{\rm div}\, u_k =0
\end{array}\right.
\end{equation}
is given by
\begin{equation}\label{e:energy-Leray}
\sup_t \left(\|u_k (\cdot, t)\|^2_{L^2} + \int_0^t \|Du_k (\cdot, \tau)\|^2_{L^2} d\tau\right) \leq M\, .
\end{equation}
In his pioneering work \cite{Leray} J. Leray used the above energy bound to prove the existence of global weak solutions to the Cauchy problem for \eqref{e:Navier-Stokes} when the initial data $u_0$ belongs to $L^2$. In the framework considered above, Leray's work can be summarized in two main discoveries:
\begin{itemize}
    \item[(i)] The space of solutions of \eqref{e:Navier-Stokes} satisfying \eqref{e:energy-Leray} is compact. The main point here is that the embedding $H^1\hookrightarrow L^2$ is compact. 
    \item[(ii)] An appropriate regularization of the equations produces a sequence of {\em approximate} solutions to the Cauchy problem with the energy bound \eqref{e:energy-Leray} and the compactness (i) can be used to show existence of am {\em exact} solution.
\end{itemize}
Of course the energy bound above is not the only possible choice; one might ask what happens if we only assume a uniform bound of kinetic energy (disregarding the cumulative dissipation). In this case the compactness argument of Leray fails, and a priori we once again merely obtain a Reynolds stress term. In a recent landmark paper \cite{BV} Buckmaster and Vicol were able to show that this is optimal in the following sense.

\begin{theorem}\label{t:Buckmaster-Vicol}
Let $(\bar{u},R)$ be a smooth solution of
\begin{equation}\label{e:Navier-Stokes-Reynolds}
\left\{
\begin{array}{ll}
&\partial_t \bar u + {\rm div}\, (\bar u\otimes \bar u + \bar R) + \nabla p = \Delta u\\ \\
&{\rm div}\, u =0
\end{array}\right.
\end{equation}
such that $R$ is symmetric and positive definite. Then there exists a sequence of weak solutions $u_k$ of \eqref{e:Navier-Stokes} such that  $u_k\rightharpoonup \bar{u}$ weakly in $L^2$ as $k\to \infty$.   
\end{theorem}

The significance of the failure of compactness in this theorem can be explained as follows. Since now \eqref{e:energy-Leray} is replaced by \eqref{e:energy1}, the only additional control on oscillatory behaviour must come from the equation itself, in particular integrability of the dissipative term. In other words, combining the inertial, Reynolds and dissipative terms in \eqref{e:Navier-Stokes-Reynolds} as one stress term of the form $(\bar u\otimes \bar u + \bar R + Du+Du^T)$, the issue is to decide whether $Du+Du^T$ dominates $\bar u\otimes \bar u$; Thus, the failure of the embedding $W^{1,1}\hookrightarrow L^2$ in 3D lies at the core of Theorem \ref{t:Buckmaster-Vicol}. 

To emphasize this point, we mention that in \cite{LuoTiti} the statement of Theorem \ref{t:Buckmaster-Vicol} was extended to weak solutions of the hyperviscous Navier-Stokes equations, where the dissipative term $\Delta u$ is replaced by $-(-\Delta u)^\theta$ with $\theta<5/4$, under the kinetic energy bound \eqref{e:energy1}. This result is related to the fact that the embedding $W^{2\theta-1,1}\hookrightarrow L^2$ is compact if and only if $\theta>5/4$. Another variation on this theme concerns power-law fluids, where $\Delta u$ is replaced by $\Delta_p u$, the $p$-Laplacian. In this case the energy bound including cumulative dissipation takes the form 
\begin{equation}\label{e:energy-Lerayp}
\sup_t \left(\|u_k (\cdot, t)\|^2_{L^2} + \int_0^t \|Du_k (\cdot, \tau)\|^p_{L^p} d\tau\right) \leq M\, .
\end{equation}
Thus, the issue of weak stability is decided by whether the embedding $W^{1,p}\hookrightarrow L^2$ is compact, and indeed: the analogue of Leray's theorem (existence of weak solutions satisfying the energy inequaliy via compactness) has been proved in \cite{Ruzicka} for $p>6/5$, whereas the analogue of Theorem \ref{t:Buckmaster-Vicol} has been proved in \cite{BSS} for $p<6/5$. 

Further interesting examples using these ideas include the optimality of the DiPerna-Lions theory for transport equations with Sobolev coefficients (see \cite{MS1,MS2,MSa,BCD}) and the optimality of the Ladyzenskaya-Prodi-Serrin regularity criterion (see \cite{CL}).

\subsection{Ideal MHD}
\label{s:MHD}

In the above examples the construction of weak solutions relied on (super-critical) scaling for the perturbations, just below the compactness range. In particular the relaxation in each case turned out to agree with the convex hull, no weakly continuous quantities remained.  
An interesting example exhibiting a relaxation which on the one hand is constrained by weakly continuous quantities but on the other hand is still sufficiently rich to be able to adapt the techniques above is provided by the ideal magnetohydrodynamic system (MHD in short). Here the incompressible Euler equations are coupled with the Faraday system via Ohm's law. The MHD system, with nonzero viscosity and magnetic resistivity, is a widely accepted model for electrically conducting fluids such as plasmas and liquid metals (see~\cite{GLBL} and~\cite{ST}). The ideal MHD system, where kinematic viscosity and magnetic diffusivity is set to zero, contains a wealth of mathematical structure \cite{ArnoldKhesinbook} and can be written as
\begin{equation}\label{e:MHD}
\begin{split}
& \partial_t u + \textrm{div }(u \otimes u-B\otimes B) + \nabla p = 0, \\
& \partial_t B + \nabla \times (B\times u) = 0,\\
& \textrm{div } u = \textrm{div }B = 0, 
  \end{split}
  \end{equation}
Moreover, in analogy with the role of the incompressible Euler equations for hydrodynamical turbulence, the ideal system is relevant in the inviscid, irresitive ``turbulent'' limit in the context of weak solutions. 

The three known conserved quantities for the ideal MHD system are the total energy $\mathcal E$, cross-helicity $\mathcal W$ and magnetic helicity $\mathcal H$, defined as
\begin{subequations}
	\begin{align*}
	\mathcal{E}(u,B)&=\frac{1}{2}\int_{\mathbb T^3}|u|^2+|B|^2\,dx,\\	
	\mathcal{W}(u,B)&=\int_{\mathbb T^3}u\cdot B\,dx,\\
	\mathcal{H}(B)&=\int_{\mathbb T^3}A\cdot B\,dx.
	\end{align*}
\end{subequations}
Here $A$ is a vector potential of $B$, so that $\textrm{curl }A=B$. In is well known and easy to verify that in the periodic setting (or in contractible domains) $\mathcal H$ is well defined and does not depend on the particular choice of potential. Thus, in analogy with the Euler equations, it is natural to study weak solutions in the energy space, that is, solutions with $u,B\in L^{\infty}(0,T;L^2(\mathbb T^3))$. In a recent remarkable paper \cite{BeekieBuckmasterVicol} the authors were able to construct, using a variant of convex integration, non-trivial weak solutions in the energy space:

\begin{theorem}
There exist weak solutions $u,B\in L^{\infty}(0,T;L^2(\mathbb T^3))$ of the ideal MHD system \eqref{e:MHD} for which none of the quantities $\mathcal{E}$, $\mathcal{W}$ or $\mathcal{H}$ are conserved in time. 
\end{theorem}

The behaviour of formally conserved quantities in classes of weak solutions is of general interest \cite{Klainerman} and of particular relevance to MHD turbulence for the following reason. Whilst anomalous dissipation (of total energy) is an observed feature of turbulent plasmas, similarly to turbulent flows, it has also been observed that the dissipation rate of total energy is order of magnitudes higher than the rate of change in total magnetic helicity, and, indeed, various astrophysical plasmas tend to evolve toward a non-trivial force-free state $\nabla \times B = \alpha B$, with magnetic helicity given by the initial configuration. A cornerstone of the Woltjer-Taylor relaxation theory \cite{Woltjer,Taylor}  is thus the expectation that total magnetic helicity remains essentially constant during the primary dissipation stage of energy, until an essentially minimal energy configuration subject to given helicity is reached. A widely accepted mechanism for this process is magnetic reconnection. On the level of mathematical analysis, one would thus expect \cite{CKS} that a physically relevant class of weak solutions of ideal MHD should reflect both the property of non-vanishing energy dissipation rate and the property of conservation of magnetic helicity. 

To obtain such a class of weak solutions, let us look again at the programme sketched above, i.e. consider a sequence of solutions $(u_k,B_k)$ of \eqref{e:MHD}, is the uniform bound
\begin{equation}\label{e:energyMHD}
    \sup_{t\leq T}(\|u_k\|_{L^2(\mathbb T^3)}+\|B_k\|_{L^2(\mathbb T^3)})\leq M
\end{equation}
for some fixed $M$. Once again we ask the question: what equation will the weak $L^2$ limit $(u_k,B_k)\rightharpoonup (\bar{u},\bar{B})$ satisfy. In other words we would like to characterize the weak closure $\overline{S}\subset \mathcal{X}=L^2\times L^2$ under the assumption that 
$\mathcal S$ consists of solutions of the ideal MHD system with the uniform bound \eqref{e:energyMHD}. 

Since  
\begin{align*}
    \partial_tB_k+\nabla \times E_k&=0,\\
    \textrm{div }B_k&=0,
\end{align*}
with electric field $E_k=B_k\times u_k$, the
uniform bound \eqref{e:energyMHD} implies also 
\begin{equation}\label{e:energyMHD2}
    \sup_{t\leq T}\|E_k\|_{L^1(\mathbb T^3)}\leq M.
\end{equation}
From this one may deduce for the associated vector potentials $A_k$ 
\begin{equation}\label{e:energyMHD3}
    \sup_k\sup_{t\leq T}(\|A_k\|_{H^1(\mathbb T^3)}+\|\partial_tA_k\|_{L^1(\mathbb T^3)})<\infty.
\end{equation}
Applying the Aubin-Lions lemma one concludes \emph{strong} convergence $A_k\to \bar{A}$ in $L^2$, and consequently that $\mathcal H(B_k)\to \mathcal H(\bar{B})$ (see \cite{FLS} and \cite{FL}, where the physically more relevant problem of the inviscid limit is treated). In particular magnetic helicity remains stable under the weak closure from $\mathcal{S}$ to $\overline{\mathcal{S}}$. 

More generally and analogously to the passage from \eqref{e:Euler} to \eqref{e:Euler-Reynolds}, we might ask for the characterization of the Reynolds-type term $R$ as well as the limiting electric field $\bar{E}$ in the system
\begin{equation}\label{e:rMHD}
\begin{split}
& \partial_t \bar{u} + \textrm{div }(\bar{u}\otimes \bar{u}-\bar{B}\otimes \bar{B}+R)+\nabla \bar{p} = 0,  \\
& \partial_t \bar{B} + \nabla \times \bar{E} = 0, \\
&   \textrm{div } \bar{u} = \textrm{div }\bar{B} = 0, 
\end{split}
\end{equation}
An interesting question in this context is to decide whether $\bar{E}\cdot \bar{B}=0$ (as compared with the original system \eqref{e:MHD}, where $E=B\times u$). Indeed, a formal calculation shows that 
$$
\frac{d}{dt} \mathcal H(\bar{B})=-2\int_{\mathbb T^3}\bar{E}\cdot \bar{B}\,dx,
$$
so that orthogonality of $\bar{B}$ and $\bar{E}$ is key to helicity conservation. This is precisely the type of question whose general study was pioneered in the work of Tartar and Murat. Indeed, we recall the following version of the div-curl lemma \cite{Tartar79, Tartar05}: 
Suppose we have a sequence of magnetic and electric fields $(B_k,E_k)\rightharpoonup (B,E)$ converging weakly in $L^p\times L^{p'}$ and such that $\{\nabla\cdot B_k\}$ and $\{\partial_tB_k+\nabla\times E_k\}$ are uniformly bounded in $L^p\times L^{p'}$. Then $B_k\cdot E_k\overset{*}{\rightharpoonup} B\cdot E$ in the space of measures. Observe that, since $E_k=B_k\times u_k$, the energy bound in \eqref{e:energyMHD} is not enough to put us in this setting. On the other hand, for instance if we assume uniform bounds on $(u_k),(B_k)$ in $L^3$, then $(E_k)$ is bounded in $L^{3/2}$ and the div-curl lemma applies. A closely related result in \cite{KL} is that in fact for any weak solution of \eqref{e:MHD} with $u,B\in L^3$ magnetic helicity is a conserved quantity. In terms of the general framework sketched in Section \ref{s:averaging} this implies that in the $L^3$ (and in particular the bounded) setting the relaxed constitutive set $K'$ has empty interior, because necessarily $\bar{B}\cdot\bar{E}=0$. Conversely, adding this condition to \eqref{e:rMHD} leads to a relaxation of ideal MHD which is consistent with the Woltjer-Taylor theory, and in particular to the following result in \cite{FLS}:

\begin{theorem}
There exist weak solutions $u,B\in L^{\infty}$ of the ideal MHD system \eqref{e:MHD} for which $\mathcal{E}$ and $\mathcal{W}$ are not conserved in time, but magnetic helicity $\mathcal{H}$ remains constant.
\end{theorem}

Finally, to close this section we recall that, beside conditions for the conservation of magnetic helicity, also the conservation of energy and cross-helicity have been studied in \cite{CKS,KL,WZ,Yu}. Conservation of the magnetic helicity was shown in~\cite{CKS} for solutions $u,B$ which are H\"older continuous in space with exponents $\theta_1,\theta_2$, respectively, with $\theta_1+2\theta_2>1$ (the statement is actually more precisely formulated in Besov spaces). This type of analysis was very much motivated by the analogous question of energy conservation for the Euler equations and Onsager's celebrated conjecture. This is the subject of the next section.

\subsection{The Onsager conjecture}

A cornerstone of turbulence theory is the K41 scaling law of the energy spectrum and, closely linked, the conjecture of Onsager on optimal H\"older exponents of dissipative weak solutions of the Euler equations. Let us recall briefly the mathematical statements.   

One interesting byproduct of \cite{DS1} is that there is a ``systematic'' way of constructing an exact (albeit weak) $L^2$ solution which approximates a triple as \eqref{e:Euler-Reynolds}. Think of $u_k$ in \eqref{e:Euler} as an exact solution of Euler $v$ which we want to be sufficiently close (in the weak topology of $L^2$) to the solution $u$ of \eqref{e:Euler-Reynolds}. Informally the paper \cite{DS1} gives an algorithm to ``eliminate'' $R$ and produce $v$ as a perturbation of $u$. 
Five years after \cite{DS1}, motivated by a celebrated conjecture by Lars Onsager, in \cite{DS3} we proposed a second algorithm which achieves the same, this time producing a {\em continuous solution}. 

Onsager suggested in his famous note \cite{Onsager} the possibility of anomalous dissipation for {\it weak solutions} of the Euler equations as a consequence of Kolmogorov's energy cascade. 
Even though the theory of Kolmogorov is a {\it statistical theory}, dealing with random fields whose distribution laws need to satisfy several postulates, Onsager stated his conjecture as a ``pure PDE'' that could be studied directly and which, after nearly 70 years, we can finally state the theorem:

\begin{theorem}\label{t:Onsager}
Let $(v,p)$ be a weak solution of \eqref{e:Euler} on the periodic $3$-dimensional torus $\mathbb T^3$ with
\begin{equation}\label{e:uniformholder}
|v(x,t)-v(y,t)|\leq C|x-y|^{\theta} \qquad \forall x,y,t
\end{equation}
(where $C$ is a constant independent of $x,y,t$). 
\begin{itemize}
\item[(a)] If $\theta> \frac{1}{3}$, then $E(t)$ is necessarily constant; 
\item[(b)] For $\theta< \frac{1}{3}$ there are solutions for which $E (t)$ is strictly decreasing. 
\end{itemize}
\end{theorem}

A first result in part (a) of Theorem \ref{t:Onsager} was fist achieved by Eyink in \cite{Eyink}. The exact statement was then proved by Constantin, E, and Titi in \cite{ConstantinETiti} using a regularization procedure and a clever and powerful, yet elementary, commutator estimate. Part (b) has been proved following a series of partial results and gradual improvements \cite{DS4,BDIS,BDS2, DaSz}, finally culminating in \cite{Isett}. In \cite{Isett} the author was able to prove, for any $\theta<1/3$, the existence of nontrivial solutions satisfying \eqref{e:uniformholder}, which are compactly supported in time. A construction, along the same lines of \cite{Isett}, of solutions which have strictly decreasing $E(t)$ has been later given in \cite{BDSV}. The regularity has been subsequently improved on a logarithmic scale in \cite{Isett3}, but the critical exponent $\theta=1/3$ remains open (for the critical result for part (a) see \cite{CCFS}).

In order to show Theorem \ref{t:Onsager} one needs to go beyond the general question of closure under natural energy bounds, and the constructions involved in the proof as well as its implications for stability and closure go well beyond the examples given in the previous paragraphs. To emphasize this point, note that the Navier-Stokes with fractional dissipation 
\begin{equation}\label{e:Navier-Stokes-fractional}
\left\{
\begin{array}{ll}
&\partial_t u + {\rm div}\, u\otimes u + \nabla p = - (-\Delta)^\alpha u\\ \\
&{\rm div}\, u =0
\end{array}\right.
\end{equation}
still has same compactness properties as the standard case, leading to weak stability of the equations and existence of Leray-Hopf solutions. At the same time the techniques leading up to Theorem \ref{t:Onsager} can be adapted to show existence (and non-uniqueness) for Leray-Hopf solutions of \eqref{e:Navier-Stokes-fractional}, cf. \cite{CDD,DeRosa}. In other words, for $\alpha <\frac{1}{3}$, solutions $(u_k, p_k)$ of \eqref{e:Navier-Stokes-fractional} under the natural ``Leray bound''
\begin{equation}\label{e:energy-Leray-fractional}
\sup_t \left(\|u_k (\cdot, t)\|^2_{L^2} + \int_0^t \|(-\Delta)^{\alpha/2} u_k (\cdot, \tau)\|^2_{L^2} d\tau\right) \leq  M\, ,
\end{equation}
enjoy weak stability (since the embedding $H^\alpha\hookrightarrow L^2$ is compact, c.f. Section \ref{s:NavierStokes}), but can also be produced using ``convex integration'' methods, a phenomenon which to our knowledge has not been observed before in any other situation. 

\medskip

While the classical justification for the exponent $\frac{1}{3}$ in Theorem \ref{t:Onsager} given in the physics literature is based essentially on dimension analysis and, crucially, the law of finite energy dissipation, the proof of Theorem \ref{t:Onsager} suggests an entirely different heuristics, which does not rely on energy considerations. 

The main mechanism to produce non-conservative solutions is in fact an algorithm which we will present in the form of an iteration scheme. Assuming that we start with a {\em sufficiently smooth}\footnote{Note that this is no serious restriction, as any triple $(v,p,R)$ can be easily smoothed to a new triple still solving \eqref{e:Euler-Reynolds}: in fact there are several ways of doing this, given that the system is very much underdetermined.} $(v_0, p_0, R_0) = (u, p, R)$ as in \eqref{e:Euler-Reynolds}, at each step of the iteration we add an oscillatory correction in order to decrease the defect to being a solution. More precisely, we construct inductively a sequence of smooth solutions $(v_q,p_q,R_q)$, $q=1,2,\dots$ to
\begin{equation}\label{e:Reynolds_2}
\left\{\begin{array}{l}
\partial_t v_q+{\rm div}(v_q\otimes v_q)+ \nabla p_q = {\rm div}\, R_q\,,\\ \\
{\rm div}\, v_q =0\,, 
\end{array}\right.
\end{equation}
such that $v_q\to v$ and $R_q\to 0$ uniformly. 

We write $v_{q+1} = v_q+w_{q+1}$ where we think of $v_q$ as the ``mean flow'' on length-scales $\geq \lambda_q^{-1}$ and $w_{q+1}$ as a ``fluctuation'' on this scale. Up to lower order corrections $w_{q+1}$ should have the form
\begin{equation}\label{e:ansatz_1}
w_{q+1}(x,t)=W\Bigl(v_q (x,t),  R_q (x,t),\lambda_{q+1} x,\lambda_{q+1} t\Bigr)\, ,
\end{equation}
where $W (v, R, \xi, \tau)$ is some ``master function'' and $\lambda_{q+1}$ a parameter which increases at least exponentially fast at each step. 

The basic idea for reducing the error with such an \emph{ansatz} is the following: assuming that $v_q$ is already the correct solution up to spatial frequencies of order $\lambda_q$, and $w_{q+1}$ is supported on spatial frequencies of order $\lambda_{q+1}$, it is easy to see that the only possibility for $w_{q+1}$ to correct the error $R_q$ is via the high-high to low interaction in the product $w_{q+1}\otimes w_{q+1}$. 
We hope to ``cancel $R_q$'' with the latter interaction and achieve a new Reynolds stress $R_{q+1}$ with Fourier support on frequencies of order $\lambda_{q+1}$ and much smaller size. In this way we can push the error to high frequencies (and reduce its size) by successively ``undoing'' the averaging process leading to Reynolds stresses. 

As explained in \cite{DS5}, starting from the \emph{ansatz} \eqref{e:ansatz_1} (and a similar one $p_{q+1} (x,t) = P (v_q (x,t),  R_q (x,t),\lambda_{q+1} x,\lambda_{q+1})$), it is possible to write down a family of ``ideal'' conditions that $W$ would have to satisfy so to give a ``clean'' iteration leading to a proof of Theorem \ref{t:Onsager}(b).
These conditions are
\begin{itemize}
\item $\xi\mapsto W(v,R,\xi,\tau)$ is $2\pi$-periodic with vanishing average, i.e.
\begin{equation}\label{e:H1}
\langle W\rangle := \frac{1}{(2\pi)^3}\int W(v,R,\xi,\tau)\,d\xi=0;
\end{equation}
\item The average stress is given by $R$, i.e. 
\begin{equation}\label{e:H2}
\langle W\otimes W\rangle = R\, .
\end{equation}
\item $W$ and $P$ satisfy
\begin{equation}\label{e:H3}
\left\{\begin{array}{l}
\partial_{\tau} W+ v\cdot\nabla_{\xi}W+{\rm div}_{\xi} (W\otimes W) + \nabla_\xi P =0\\ \\
{\rm div}_\xi W = 0\, .
\end{array}\right.
\end{equation} 
\end{itemize}
Assuming a $W$ as above exists and it is smooth in all variables\footnote{In reality this family of conditions is somewhat naive and there is no such $W$ which would satisfy the bound \eqref{e:H4}. The main issue is in the advective part $\partial_\tau v_q + v\cdot \nabla_\xi W$ of equation \eqref{e:H3}. While it is not possible to accomodate \eqref{e:H3} exactly, the main idea around the problem is that the condition does not need to be fulfilled ``exactly'', but we can allow for a suitable error. In a nutshell all the literature leading from \cite{DS3} to \cite{Isett} (and its improvement \cite{BDSV}) has been an effort of making the latter error as tame as possible.}, the best we can hope, in view of \eqref{e:H2} are bounds of the form
\begin{equation}\label{e:H4}
|W|\lesssim |R|^{1/2},\,|\partial_vW|\lesssim |R|^{1/2},\,|\partial_R W|\lesssim |R|^{-1/2}.
\end{equation}
Assuming the existence of a such a profile $W$, the next stress tensor $R_{q+1}$ 
would then be defined through
\begin{eqnarray}
R_{q+1}&=\;-&{\rm div}^{-1}\Bigl[\partial_tv_{q+1} +{\rm div}\, (v_{q+1} \otimes v_{q+1})+\nabla p_{q+1}\Bigr]\notag\\
&=\;-&\underbrace{{\rm div}^{-1}\Bigl[\partial_t w_{q+1} +v_q \cdot\nabla w_{q+1} + {\rm div}\, (w_{q+1}\otimes w_{q+1}- R_q)+\nabla (p_{q+1}-p_q)\Bigr]}_{=: R_{q+1}^{(1)}}\nonumber\\
&\phantom{=}\; &- \underbrace{{\rm div}^{-1}\Bigl[w_{q+1}\cdot \nabla v_q\Bigr]}_{=:R_{q+1}^{(2)}}
\end{eqnarray}
where ${\rm div}^{-1}$ is a suitable operator of order $-1$ which inverts the divergence. 

In order to gain an understanding of the size of the two terms $R_{q+1}^{(i)}$ we can expand the argument of the operator ${\rm div}^{-1}$ in Fourier series in the fast variables, with coefficients which depend on the slow variables. We illustrate this principle in the case of for the second term:
\begin{equation}\label{e:estR3}
R^{(2)}_{q+1} ={\rm div}^{-1}\Bigl[w_{q+1}\cdot \nabla v_q\Bigr]={\rm div}^{-1}\sum_{k\in\mathbb Z^3, k\neq 0}c_k(x,t)e^{i\lambda_{q+1} k\cdot x}\, .
\end{equation}
The coefficients $c_k (x,t)$ vary much slower than the rapidly oscillating exponentials and moreover the coefficient $c_0$ vanishes: this is in fact the content of condition \eqref{e:H1}. When we apply the operator ${\rm div}^{-1}$ we can therefore treat the $c_k$ as constants and gain a factor $\frac{1}{\lambda_{q+1}}$ in the outcome: a typically ``stationary phase argument''. 
The conditions \eqref{e:H1} and \eqref{e:H2} ensure that the same argument can be applied when dealing with $R_{q+1}^{(1)}$. \eqref{e:H1} ensures that when the differential operators hit the fast exponentials the resulting term vanishes, while \eqref{e:H2} guarantees that the Fourier expansion in the fast variables misses the zero frequency coefficient: \eqref{e:H2} encodes the high-high to low interaction with which the perturbation ``cancels'' $R_q$.  

Returning to \eqref{e:estR3} observe that:
\begin{itemize}
\item If $\delta_{q+1}^{1/2}$ is the size of the perturbation $w_{q+1}$, the condition \eqref{e:H2} implies that the size of $R_q$ must be $\delta_{q+1}$;
\item Since the frequency of the perturbation $w_q$ is $\lambda_q$, the size of $Dv_q$ is $\delta_q^{1/2} \lambda_q$;
\item We thus expect the size of $R_{q+1}^{(2)}$ to be $\delta_{q+1}^{1/2} \delta_q^{1/2} \lambda_q \lambda_{q+1}^{-1}$.
\end{itemize}
We do not enter into similar considerations for the term $R_{q+1}^{(1)}$: it can be readily checked that, under the above conditions, we can expect it to be of smaller size. Thus, in order to close the iteration scheme we need the relation
\begin{equation}\label{e:recurrent}
\delta_q^{1/2} \delta_{q+1}^{1/2} \frac{\lambda_q}{\lambda_{q+1}} \leq \delta_{q+2}\, .
\end{equation}
Assuming an exponential growth of the frequencies $\lambda_q = \lambda^q$ and a related exponential decay of the sizes $\delta_q = \lambda^{-2\alpha q}$, it can be readily checked that \eqref{e:recurrent} leads to the constraint
\[
- \alpha q - \alpha (q+1) +q - (q+1) \leq - 2\alpha (q+2)
\]
i.e. $\alpha \leq \frac{1}{3}$. Since the $C^\beta$ size of the perturbation is $\delta_{q+1}^{1/2} \lambda_q^\beta$, the critical threshold $\frac{1}{3}$ appears naturally for the H\"older convergence of the sequence.

\subsection{The inviscid limit}

Recalling that Kolmogorov's theory pertains to the Navier-Stokes equations, a very appealing question is: can we produce a sequence of solutions to $(u_q, p_q)$ of 
\begin{equation}\label{e:Navier-Stokes-vanishing}
\left\{
\begin{array}{ll}
&\partial_t u_q + {\rm div}\, u_q\otimes u_q + \nabla p_q = \nu_q \Delta u_q\\ \\
&{\rm div}\, u_q =0
\end{array}\right.
\end{equation}
with viscosity $\nu_q$ converging to $0$, bounded kinetic energy and which converge to a dissipative weak solution $(u,p)$ of the Euler equations? 
As pointed out in \cite{CG,CV,DE}, assuming a uniform energy spectrum as in Kolmogorov, this is to be expected. 
A tempting conjecture would be that at least some of the weak solutions of the Euler equations produced via the methods described in the previous paragraphs can be approximated with solutions of the Navier-Stokes equations. 

Remarkably, at the level of weak solutions of Navier-Stokes, this has been proved by Buckmaster and Vicol in \cite{BV}. 
However, the solutions produced in \cite{BV} are highly irregular and, more importantly, do not satisfy the stronger condition ``\`a la Leray'' \eqref{e:energy-Leray}.
A totally open question is thus whether any of the chaotic behavior witnessed by weak solutions of the Euler equations can in fact be achieved by limits of {\em Leray} solutions of the Navier-Stokes equations with bounded kinetic energy. 

One possible way to phrase the latter question is the following. Consider the right hand sides ${\rm div}\, R_q$ in the equation \eqref{e:Reynolds_2} produced by the iteration scheme described in the previous paragraph. How far is such ${\rm div}\, R_q$ from $\nu_q \Delta v_q$? Recall that the solutions produced by the scheme are in fact smooth, and so if we were able to produce them so that ${\rm div}\, R_q=\nu_q \Delta v_q$, we would in fact get a sequence of {\em smooth} solutions of Navier-Stokes. 

At the qualitative level there are some remarkable similarities between the terms ${\rm div}\, R_q$ generated by the proof of Theorem \ref{t:Onsager} and what the K41 theory predicts for the viscous dissipation $\nu_q \Delta u_q$.
A first elementary observation is that $\nu_q \Delta v_q = {\rm div}\, \nu_q (Dv_q+ Dv_q^T)$ and that $\nu_q (Dv_q+Dv_q^T)$ is a traceless symmetric tensor. These two properties can be shown to hold for the tensors $R_q$ produced by pretty much all schemes following the one of \cite{DS3}.

A second observation is that, as already explained in the previous paragraph, along the iteration leading to the proof of Theorem \ref{t:Onsager} the $L^\infty$ sizes of $R_q$ and $v_q$ are $\delta_{q+1}$ and $\delta_q^{1/2}$, while their typical length scale is $\lambda_q^{-1}$. Recalling that the Onsager threshold is given by $\delta_q= \lambda_q^{-2/3}$, the size of $Dv_q$ is typically $\lambda_q^{2/3}$. Since $\frac{\delta_{q+1}}{\delta_q}$ is expected to be constant, the size of $R_q$ is $\lambda_q^{-2/3}$ and we conclude that the relation $R_q = \nu_q (Du_q+Du_q^T)$ would require $\nu_q = \lambda_{q}^{-4/3}$. In other words, given that $\ell = \lambda_{q+1}^{-1/3}$ is the typical length of the latest correction, $\nu_q$ would have to be the Kolmogorov's length scale. 
Thus the proof of Theorem \ref{t:Onsager} leads to a sequence of approximations of a dissipative solution of Euler which can be interpreted as solutions to an ``artificial viscosity'' regularization: with the latter interpretation, the scaling of this artificial viscosity is indeed the one predicted by Kolmogorov's theory\footnote{This is in fact not fully correct since the proof of Theorem \ref{t:Onsager} does not achieve the threshold $\frac{1}{3}$ but rather gets $\varepsilon$ close to it for every positive $\varepsilon$.}.

Clearly, such artificial viscosity misses much of the structure of the actual viscosity in \eqref{e:Navier-Stokes-vanishing}. For instance it misses the estimate \eqref{e:energy-Leray}. In fact smooth solutions of \eqref{e:Navier-Stokes-vanishing} satisfy the stronger identity
\begin{equation}
\partial_t \frac{|u_q|^2}{2} + {\rm div}\, \left(\left(\frac{|u_q|^2}{2} + p\right) u_q \right) + \nu_q |Du_q|^2 = \nu_q \Delta \frac{|u_q|^2}{2}\, .
\end{equation}
As remarked in \cite{DuchonRobert}, if $u_q$ converge strongly in $L^3$ to some $u$, then the latter would be a weak solution of incompressible Euler which satisfies the local energy inequality
\begin{equation}\label{e:energy-dissipation-local}
\partial_t \frac{|u|^2}{2} + {\rm div}\, \left(\left(\frac{|u|^2}{2} + p \right)u\right) \leq 0\, .
\end{equation}
This suggests a stronger form of the Onsager conjecture, where the dissipative solution is required to satisfy \eqref{e:energy-dissipation-local}. The first such examples in the literature were given in \cite{DS2}, but the solutions in \cite{DS2} were only bounded. H\"older weak solutions were first constructed by Isett in \cite{Isett2} and the H\"older regularity found in \cite{Isett2} has been later improved in \cite{DK}, but these results are still relatively far from the $\frac{1}{3}$ threshold.

\end{document}